\documentclass[twocolumn]{esapub}
\usepackage{graphicx,times,natbib}

\title{The Magnetic Field in the Convection Zone.
}
\author{Alberto Bigazzi}
\author{Alexander A. Ruzmaikin}
\affil{Jet Propulsion Laboratory, California Institute of Technology, CA USA}
\date{\today}
\begin{document}
\maketitle
\abstract{
One of the key questions in solar physics that remains to be answered 
concerns the strength and the distribution of the magnetic fields
at the base of the convection zone. 
The flux tube dynamics requires that toroidal fields of strength as large as
100 kilogauss be present at the base of the convection zone. 
The kinetic-magnetic equipartition argument leads to smaller field 
strengths. For possible detection of these relatively small (compared 
to pressure effects) fields by helioseismic methods it is important 
to know the range of the field strengths and their distribution.

We estimate a range for the toroidal magnetic field strengths at the 
base of the convection zone using dynamo simulations in a spherical 
shell. These  simulations involve the distribution of 
rotation provided by helioseismic inversions of the GONG and MDI 
data. Combining the simulations with the observed line-of-sight 
surface poloidal field we extract the spatial pattern and magnitude 
of the mean toroidal magnetic field at the base of the convection 
zone. 
}
\section{Introduction}
The inner distribution  of the solar magnetic field is poorly known. 
Despite the striking regularity of the solar activity, we still lack an
understanding of its causes and the mechanisms by which it operates. 

The layer that we are able to directly observe the magnetic fields 
on is the photosphere. 
From observations of sunspots we know that magnetic fields with
intensities of
the order of thousands of Gauss are produced somewhere  below the photosphere 
and can live for months before decaying away. 
The topological and  magnetic structure of sunspots (as expressed by Hale's 
and Joy's laws) 
suggest that they harbor intense, toroidal magnetic flux tubes. 
The poloidal component of the solar magnetic field is much weaker compared
with the toroidal field. 

The intensity and spatio-temporal distribution of magnetic fields inside the
Sun represent a challenge  for any theory explaining the origin of solar
activity.
On one hand, the highly localised  sunspot field  points toward 
local dynamics. 
On the other, the much more diffuse poloidal field more naturally fits
into a mean-field model. 
One needs to combine  two approaches to explain the solar cycle:
the flux tube dynamics and the mean field dynamo. 
Thin flux tube dynamics has been very successful in explaining 
features connected to sunspots, see e.g Choudhuri \& Gilman 1987,  
Fan Fisher \& De Luca 1993, Caligari, Moreno-insertis, \& Sch\"ussler 1995.
The periodic migration of
sunspots toward the equator and the field reversal cannot be explained by this
dynamics alone. 
The explanation of the reversals of the field and of the equator-ward migration
of the activity were in fact one of the great achievements of the mean field
dynamo theory originated by Parker, Steenbeck, Krause and R\"adler (Parker 1955,
Steenbeck, Krause and R\"adler 1969).
Yoshimura (1975a) proved that waves mainly  propagate  along the
isosurfaces of angular velocity. 
With the present knowledge of solar differential rotation, this would imply
that inside the bulk of the convection zone waves travel radially outwards, not
giving rise to equatorward migration. 
Migration may be restored in the shear layer at the base of the convection
zone.

While it is possible to generate high magnetic fluxes by means of differential
rotation, to store them for enough time to allow for their intensity to build
up would not be feasible in the convection zone  because 
such magnetic flux tubes would erupt in a timescale of months (Parker 1975, 
Moreno-insertis 1986). 
Helioseismology tells us that around 0.7~$\mbox{R}_{\odot}$ , a sharp radial change in
the solar rotation curve happens in a layer whose thickness can be as small as
0.02~$\mbox{R}_{\odot}$ (see  Christensen-Dalsgaard et al. 1991, Basu \&
Antia 1997, Kosovichev 1996, Corbard et al. 1998, Charbonneau et al. 1999.
). 

 The same layer may allow
for fields up to $10^5$  Gauss to be stored, what would be needed  
for thin flux tube dynamics to work in the case of the Sun (Moreno-Insertis
et al. 1992, Ferriz-Mas \& Sch\"ussler 1994) 

Following Ivanova \& Ruzmaikin (1977), Parker (1993) discussed a model of mean field dynamo that includes a sharp gradient of turbulent diffusivity and 
two distinct location for the sources of the
magnetic field, that is differential rotation and helical turbulent motions. 
Stronger toroidal fields can in fact 
 be produced in the region where diffusivity is
smaller, just below the convection zone. 
Separating the shear layer and the source of the alpha-effect, moreover,
would allow for alpha not be quenched by the strong underlying magnetic
field. 
A very thorough investigation of this kind of models has been carried out by 
Charbonneau and MacGregor, 1997. 
We will use in our Profile II a similar setup to one of those discussed in
the aforementioned paper. 

In the following, we are going to study 
how different profiles of the $\alpha$-effect may influence the spatial
distribution of fields. 
We shall consider the profile of diffusivity and the rotation curve as
given.  

Mean field, kinematic dynamo cannot predict the absolute values of the
generated magnetic fields, only their ratio. 
The measured mean radial field at the surface can  then be used to 
infer the value of the toroidal magnetic field in the interior, once the
ratio is known. 

\section{Dynamo Model}
\begin{figure*}[t!]
\centering
\includegraphics[width=0.8\linewidth]{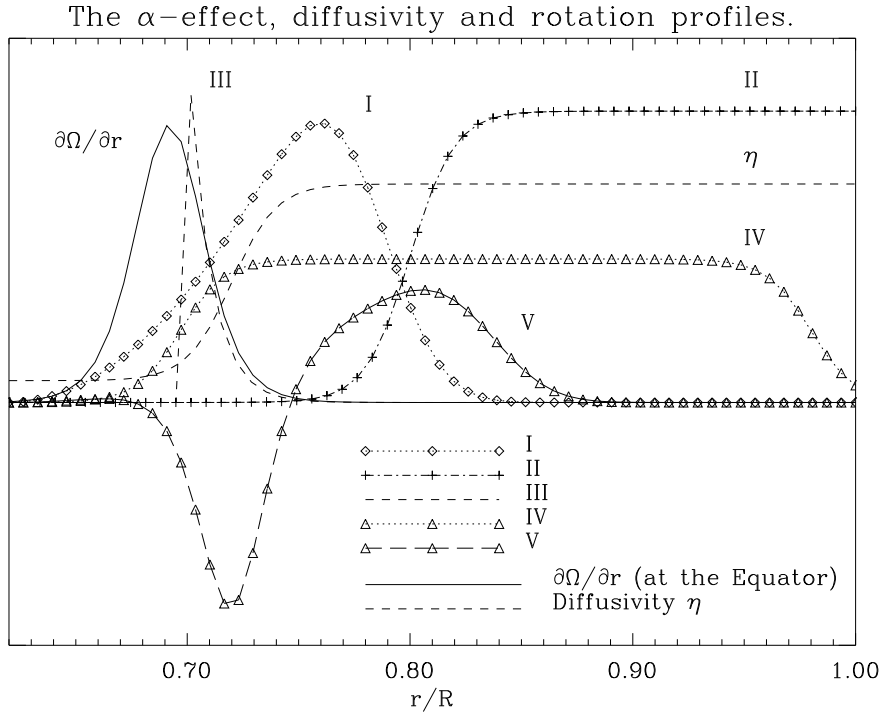}
\caption{
Radial distribution of the $\alpha$-effect for the cases I-V. 
Values are not to scale. See Table~\ref{T_alp} for the actual values used 
in the present simulations.  
The solid lines, dashed and continuous, represent the radial profile of the turbulent magnetic diffusivity $\eta$ and
the radial gradient of the rotation rate $\partial\Omega / \partial r$ at the
equator, respectively. 
They 
define for our model the Convection Zone and the shear layer ( the Tachocline). 
}
\label{F-profiles}
\end{figure*}

We consider axisymmetric solutions of the mean field dynamo. 
We assume for the solar rotation a simple analytical fit to the profile 
reconstructed by helioseismic where the surface rotation curve 
\begin{equation}
\Omega_{\mbox{s}}=\Omega_{\mbox{Eq}}\left(
1+a_1\cos^2\theta+a_2\cos^4\theta
\right)
\end{equation}
is made smoothly  match the core rotation  $\Omega_{\mbox{c}}$ in a layer of
thickness $0.2 \mbox{R}_{\odot}$ at $0.692 \mbox{R}_{\odot}$.
Equatorial rotation is $\Omega_{\mbox{Eq}}=2.865\times10^-6 \mbox{s}^{-1}$
and the core rotation taken as the value of
$\Omega_s $ at $30^{\circ}$ latitude. $\theta$ here is colatitude. 
$ a_1=-.126 $ and $ a_2=-0.159 $.
The radial profile of $\partial \Omega /\partial r $ at the equator is shown
in solid line in  Figure~\ref{F-profiles}. The sign of this gradient is
opposite at higher and lower latitudes. 
Turbulent diffusivity $\eta$ is constant throughout the Convection Zone and
we have it drop a factor $10$ to $50$, in a layer of thickness $0.2 \mbox{R}_{\odot}$ at a
location either coincident with that of the rotational shear layer or slightly
above it, at $0.713 \mbox{R}_{\odot}$. 
Both these values have been worked out in
the context of Helioseismology, see Christensen-Dalsgaard et al. 1991, Basu \&
Antia 1997, Kosovichev 1996, Corbard et al. 1998, Charbonneau et al. 1999. 
In reality the drop in turbulent diffusivity is estimated to be of the order
of $10^6$.
In Figure~\ref{F-profiles}
the diffusivity profile along the radial direction is represented
 by the solid dashed 
line, its gradient marking the bottom of the Convection zone. In solid line, 
the radial derivative of the rotation which defines the shear layer is
plotted.  

The mean field dynamo equation, written in terms of scalar potentials (Krause \& R\"adler,
1980) is solved numerically in a $\mbox{r}-\theta$ meridional 
semi-disk. Second order finite differences in space and a third order
Runge-Kutta scheme for time advance are used. In most of the runs a grid of
$60\times 40$ is used. 
Ideal conductor boundary conditions are used at the interface with the core and 
radial field condition is implemented at the surface.  Regualrity condition
are imposed on the axis. 

No model for non-linear $\alpha$-quenching has been used. This is consistent
with our assumption that the flow field is not influenced by the magnetic
field.

\subsection{The $\alpha$-effect}

Given the lack of constraints on the form of the $\alpha$-effect, 
we shall discuss how different alphas influence the spatial distribution of the 
the toroidal and poloidal fields.
Figure~\ref{F-profiles} shows the distribution in radius of $\alpha$. 
In all models, a latitudinal dependence of $\cos(\theta)$ is  included. 
This is standard in dynamo theory and reflects the property of the
alpha-effect of being antisymmetric with respect to the Equator. 
In case IV, an additional factor of $\sin^2(\theta)$ is present. This has 
been discussed by R\"udiger \& Brandenburg, 1995, and has the
effect of shifting the magnetic field patterns closer to the equator, which
better reproduces the observed patterns of sunspot migration.

Profile I, the diamonds in Figure~\ref{F-profiles}, is maximum in the bulk of
the Convection Zone, at $0.82$~$ R_{\odot}$ in our case. It is null  at
the surface.  This form of the $\alpha$-effect
takes into account the influence o the rotation on helical turbulence, see
Zeldovic, V. Ruzmaikin, Sokoloff 1983. 
Profile II, squares, is sharply peaked at $0.7$~$ R_{\odot}$, just above the shear layer, and represents an  Interface Dynamo model. As already mentioned, 
this profile was used by Charbonneau and MacGregor (1997). 
Profile III is, instead, non vanishing only in the outer shell of the Convection
Zone. This kind of ``surface dynamo'' has been discussed  in conjunction with
meridional circulation, see e.g. Choudhouri et al. 1995.
Profile IV, triangles, has a constant value throughout the whole
Convection Zone. It drops to zero close to the surface and at the base of
Convection Zone. This profile has often be adopted in the literature. 
The last case considered, V, $\alpha$ has two contributions: 
one coming from the bottom of the
convection zone where it is negative and proportional to the  gradient of
$\eta$,  combining  the effects due to the  the decrease in the intensity 
of the turbulence and the stratification. The other, coming from the
convection zone, is taken to be the same as in I. 
Yoshimura (1975b) has used this form for the $\alpha$-effect.

\begin{table}[h]
\begin{tabular}{|l|r|r|r|r|r|r|}
\hline
$\alpha (r)$& $r_{\alpha}$ 	& $\alpha_0$ & 
$\gamma$ & $T$
              & $r_{\mbox{max}}$ & $B_{\mbox{t}}/ B_{\mbox{r}}$ \\
\hline
I & $.75$ 	& 4.1 		& 8.2 & .026 & .71, .84 &  310 \\
II & .7  		& 27.5  	& 19 & .047  & .67, .80 & 950 \\
III & $>.8$		& 4.1		& 18 &  .008 & .78, .92&36  \\
IV & $.7 \div .98$	& 4.23		& 5  & .015  & .9     & 73  \\
V & $.75$		& 5		& 180  & .020 & .71   & 360  \\
\hline
\end{tabular}
\caption{ 
Location and intensity of the maximum toroidal magnetic field for
different choices of the profile of the $\alpha$-effect.  
The location where the profile is peaked $r_{\alpha}$ is given, along with the
intensity  $\alpha_0$. $\gamma$ and $T$ are the growth rate and the period 
of the solutions.  
$B_{\mbox{r}}$ is
the maximum value of the radial photospheric field averaged
over time. $B_{\mbox{t}}$ is the maximum value of the time average of 
the  toroidal magnetic field intensity.   When two values are given, the
first one 
represents a secondary maximum close to the shear layer, see
Figure~\ref{F_TPRuz}.
$r_{\mbox{max}}$ is the location where the maximum is located, in units of $R_{\odot}$.  
Numbers  are adimensional. Lengths have been scaled to $R_{\odot}$ and time
to $R_{\odot}^2/\eta$, where $\eta$ is the value of turbulent diffusivity in
the convection zone.
}
\label{T_alp}
\end{table}

\begin{figure*}[ht!]
\centering
\includegraphics[width=0.8\linewidth]{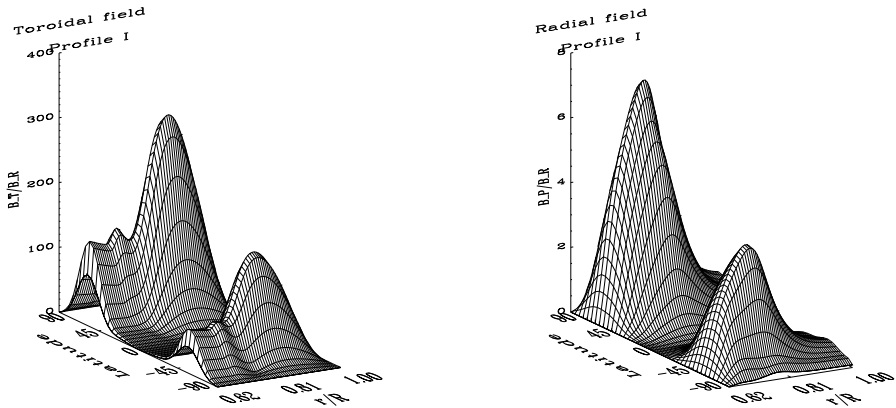}
\caption{
Surface plot of toroidal and radial field intensity 
integrated over time for Profile I. Values are scaled to the maximum of the
radial surface field.  
}
\label{F_TPRuz}
\end{figure*}

\begin{figure*}[ht!]
\centering
\includegraphics[width=0.8\linewidth]{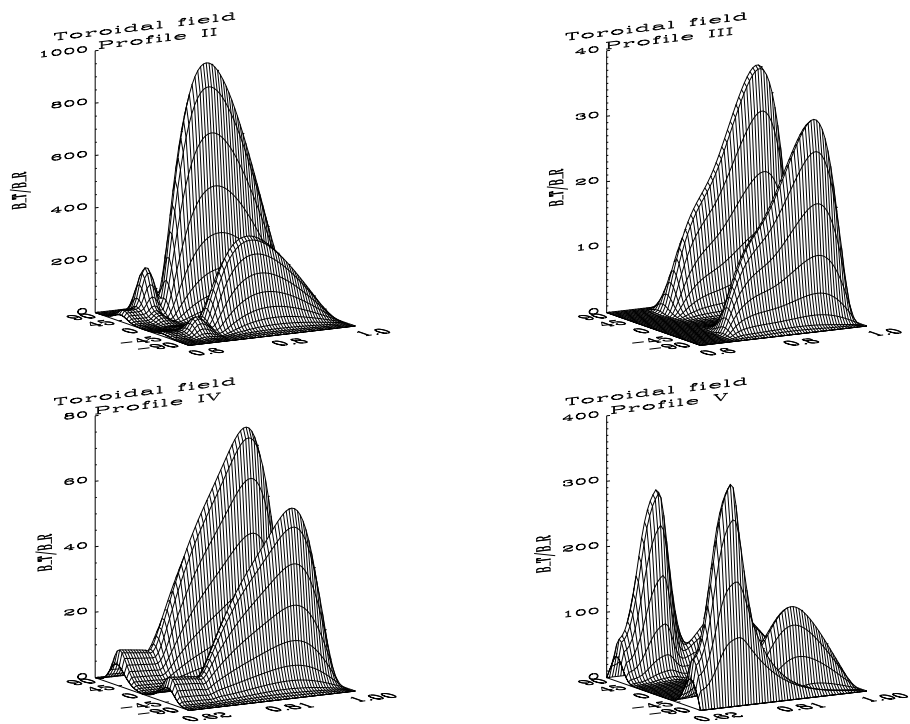}
\caption{
Toroidal fields for Profile II-V. Values are scaled to the maximum of the
radial surface field.
}
\label{F_Tall}
\end{figure*}
\section{Discussion of the results}
Except for the case of Profile III, where $\alpha$ is concentrated in 
the surface layer, all the case studied display a local maximum of the
toroidal field in the shear layer. Except for case V, this is not an absolute
maximum which is instead achieved in the convection zone, within
$0.8\div0.9$~$ \mbox{R}_{\odot}$, see Figure~\ref{F_TPRuz} and 
Figure~\ref{F_Tall}.
 
The ratio of the maximum toroidal field to the surface radial field can vary
from a few tens as in the case of the surface $\alpha$-effect, case III, 
to a thousand, as in the case of the interface profile II. 
Both the profile I and V have ratios of the order of a few hundreds. 
No symmetry across the equator has been imposed on the solution which display
a North-South asymmetry. Except for case of the surface $\alpha$, the
symmetry of the solution is mainly dipolar. 
Considering a mean radial surface field of the order of a Gauss
(Schlichenmaier and Stix 1995) the range of ratios that we found would lead
to an estimate of the large-scale mean toroidal field of the order 
of $10^3$ Gauss in the most favorable cases I,II,V. 
Both the surface (III) and top-hat (IV) profiles have smaller ratios of the
toroidal to the surface fields. 

If one assumes that solar activity is originated in the shear layer, then it
is possible to draw a \emph{butterfly diagram} with the temporal evolution of
the toroidal field near the shear layer. 
All those models show butterfly diagrams with activity at higher latitude
then observed. This is a known feature of many dynamo models. 

We have shown that different assumptions about the $\alpha$-effect give rise
to very different spatial distributions of the magnetic fields inside the
Sun. This messsage could also be read in reverse: should we be able to
probe the field deeply inside the Sun, we could have information about the
nature of the regeneration mechanism expressed by the $\alpha$-effect.

\section*{Acknowledgments}
This work was performed while A. Bigazzi held a National Research Council
Research Associateship Award at the Jet Propulsion Laboratory. 
This research was conducted in part at the Jet Propulsion Laboratory, California
Institute of Technology, under contract with NASA. 
{}

\begin{thebibliography}{}
\bibitem{}
Fan, Y., Fisher, G. H., Deluca, E. E., 1993, ApJ 405, 390
\bibitem{}
Choudhuri, A.R., Gilman, P. A., 1987, ApJ 316, 788
\bibitem{}
Parker E.N., 1975, Apj 198, 205
\bibitem{}
Moreno-Insertis F., 1986, A\&A 166, 291
\bibitem{}
Moreno-Insertis, F., Sch\"ussler M., Ferriz-Mas, A., 1992, A\&A, 264 686
\bibitem{}
Ferriz-Mas A.,  Sch\"ussler M.,  1994, ApJ, 433, 852
\bibitem{}
Parker, E. N., 1955, ApJ. 122, 293 
\bibitem{}
Steenbeck,M., Krause F., R\"adler K.-H., 1966, Z. Naturforsch., 21a, 369
\bibitem{}
Parker, E.N., 1993 ApJ 408, 707
\bibitem{}
Ivanova, T.S., Ruzmaikin A., 1977, Sov. Astronom. 21, 479
\bibitem{}
Krause F.,  R\"adler K.-H., 1980, Mean-field Magnetohydrodynamics and dynamo
theory, Akademic Verlag, Berlin. 
\bibitem{}
R\"udiger G., Brandenburg A., 1995, A\&A, 296,557
\bibitem{}
Yoshimura H., 1975a, ApJ 201, 740 
\bibitem{}
Yoshimura H., 1975b, ApJs 294, 29, 467 
\bibitem{}
Zeldovic Ya.B., Ruzmaikin A.A., Sokoloff D.D., 1983 Magnetic Fields in
Astrophysics, Gordon and Breach. 
\bibitem{}
Choudhuri, A.R. Sch\"ussler, M., Dikpati M., 1995, A\&A, 303, L29
\bibitem{}
Charbonneau P., MacGregor K.B., 1997, ApJ 486, 502
\bibitem{}
Caligari P., Moreno-Insertis F.,  Sch\"ussler M., ApJ 441, 886
\bibitem{}
Christensen-Dalsgaard J. Gough D.O., Thompson M.J., 1991, ApJ 378, 413 
\bibitem{}
Basu S., Antia H., 1997, MNRAS 287,189
\bibitem{}
Kosovichev A.G., 1996, ApJ 469 L61
\bibitem{}
Corbard, T., Berthomieu, G., Provost, J., Morel, P., 1998, A\&A, 330, 1149
\bibitem{}
Charbonneau, P., Christensen-Dalsgaard, J., Henning, R. et al., 1999, ApJ
527, 445
\bibitem{}
Schlichenmaier R., Stix M., 1995, A\&A 302, 264
\end{thebibliography}
\end{document}